\documentclass[11pt]{article}
\usepackage{hyperref}
\pdfoutput=1

\begin{document}

\title{High speed shadowgrpah of a L/D cavity at Mach 0.7 and 1.5}
\author{Ryan Schmit, Frank Semmelmayer, Mitch Haverkamp and James Grove\\ 
\\\vspace{6pt} Air Force Research Laboratory, \\ Wright-Patterson Air Force Base, OH 45433, USA}

\maketitle

\begin{abstract}
This is article highlights the fluid dynamics video of a rectangular cavity with an L/D of 5.67 at Mach 0.7 and 1.5.
\end{abstract}

A rectangular cavity with an L/D of 5.67 at Mach 0.7 and 1.5, Reynolds number $2x10^6$ and $2.3x10^6$ respectively, was examined using high speed shadowgraph imaging.  The cavity motion is shown in the \href{http://ecommons.libratry.cornell.edu/bitstream/1813/8237/2/LargeCavityMovie.mpg}{video}. The three movies clips presented were sampled at at 75kHz and played back at 20 Hz.  The camera shutter-speed was 0.37$\mu$sec.  The first clip shows the side and top view of the cavity at Mach 0.7.  Note that these two views are not in sync.  The second clip shows the side and top view of the cavity at Mach 1.5.  Again the views are not in sync.  The third clip shows a zoomed out side view of the cavity at Mach 1.5.  \\

For more information please refer to Schmit, Semmelmayer, Haverkamp and Grove, "Fourier Analysis of High Speed Shadowgraph Images around a Mach 1.5 Cavity Flow Field", \it{29th AIAA Applied Aerodyanmic Conference}, Honolulu, Hi, pp 1-24, 2011, AIAA 2011-3961

\end{document}